\documentclass[%
 reprint,
 amsmath,amssymb,
 aps,
]{revtex4-2}

\usepackage{graphicx}
\usepackage{dcolumn}
\usepackage{bm}
\usepackage{color}

\begin{document}

\title{Thermalization and irreversibility of an isolated quantum system II}

\author{Xue-Yi Guo}
\email{guoxy@baqis.ac.cn}
\affiliation{
 Beijing Academy of Quantum Information Sciences, Beijing 100193, China
}%

\date{\today}

\begin{abstract}
The irreversible entropy increase described by the second law of thermodynamics is fundamentally tied to thermalization and the emergence of equilibrium. In the first part of our work \citep{guo_thermalization_2025}, we constructed an isolated gas system model and numerically demonstrated irreversible growth of entanglement entropy caused by erasure of spread non-equilibrium state information.
Here, we mathematically prove that for a typical macroscopic system in any non-equilibrium state $|\phi_0\rangle$, the quantum state $|\phi'_0\rangle = \hat{O}(t)|\phi_0\rangle$ will inevitably evolve toward equilibrium. Our work demonstrates that the second law of thermodynamics—and consequently the ergodic hypothesis in statistical physics—can be understood and proven from a quantum information perspective.
From this perspective, the second law can be stated as: In typical macroscopic physical systems, the spreading and erasure of non-equilibrium state information is inevitable.
\end{abstract}

\maketitle

The equilibrium state of an isolated macroscopic system represents a special class of states characterized by the following fundamental features. From a macroscopic perspective, the system's observable mechanical quantities remain time-invariant. Microscopically, the quantum states corresponding to equilibrium satisfy Boltzmann's ergodic hypothesis or Gibbs' equal-a-priori-probability postulate - that all accessible microstates are equally likely.
Moreover, equilibrium states exhibit remarkable stability: local perturbations, operations, or even measurements cannot disrupt their equilibrium nature. These equilibrium states and their ergodic properties form the foundation of statistical physics. Despite their special characteristics, such states are ubiquitous and may even be considered inevitable. This inevitability stems from the second law of thermodynamics, which dictates that the entropy of an isolated system cannot decrease. The implicit meaning here is that the entropy of an isolated system can increase, and this entropy-increasing process is irreversible.
Consequently, when two macroscopic systems at different temperatures come into contact, energy and particles will transfer and redistribute between them, driving the systems continuously toward equilibrium. From this perspective, the irreversible entropy increase described by the second law of thermodynamics serves as the fundamental prerequisite for the existence of equilibrium states and their ergodic hypothesis.

A fundamental question regarding equilibrium states is whether states with such properties truly exist. The Eigenstate Thermalization Hypothesis (ETH) (see reviews \citep{deutsch_eigenstate_2018, dalessio_quantum_2016}) states that for a system with complex internal interactions (non-integrable system), its energy eigenstate $|\phi\rangle_E$ (non-degenerate due to interactions) is an equal-probability superposition of the energy eigenstates $|\phi_0 \rangle_{E,i}$ (degenerate states) of the corresponding non-interacting system (integrable system). Consequently, these energy eigenstates $|\phi\rangle_E$ exhibit properties of thermal equilibrium states—remaining time-independent with well-defined energies—which is also consistent with the isolation requirements of the ergodic hypothesis.
The ETH characterizes and distinguishes integrable from non-integrable systems through their energy level spacing statistics. Non-integrable systems, due to interaction-induced level repulsion, satisfy the Wigner-Dyson distribution.
However, what specific forms of internal interactions in isolated system Hamiltonians meet these requirements? Numerous studies have focused on transforming various many-body system models from integrable to non-integrable (see reviews \citep{deutsch_eigenstate_2018, dalessio_quantum_2016}).
It should be noted that early ETH research neglected the study of the internal entanglement properties of physical systems
and dynamical processes of entanglement growth.
Recent works \citep{wilming_entanglement-ergodic_2019, bianchi_typical_2019, huang_universal_2019, lydzba_eigenstate_2020, huang_universal_2021, lydzba_entanglement_2021, bianchi_volume-law_2022} have investigated entanglement properties of Many-body system's energy eigenstates. However, research on the dynamical process of entanglement growth and its irreversibility remains lacking.
Thermalization is a dynamical evolution phenomenon where a many-body system transitions from non-equilibrium to equilibrium states, involving both energy/particle transport and redistribution, as well as entanglement generation and spreading among initially non-equilibrium subsystems after their interactions.
This entanglement dynamics constitutes the core phenomenon of thermalization. Therefore, we expect our work to serve as a valuable complement to ETH research.

Given the ambiguity surrounding non-integrable systems in the Eigenstate Thermalization Hypothesis (ETH), research \citep{chamon_emergent_2014} suggests that investigating the existence of thermal equilibrium states does not require analysis of energy levels and their spacing statistics. Instead, one can focus solely on the properties of thermal equilibrium states themselves.
To this end, their study employed random quantum circuits to prepare maximally entangled states, then used disentangling algorithms to determine whether these states could be efficiently disentangled - thereby verifying if they represent genuine thermal equilibrium states satisfying the second law of thermodynamics. The difficulty of disentangling these maximally entangled states is quantified by two metrics: \textit{entanglement complexity} \citep{yang_entanglement_2017} and \textit{circuit complexity} \citep{chamon_circuit_2024}. These studies demonstrated that the statistical properties of entanglement spectra can characterize the entanglement complexity of thermal equilibrium states.
Furthermore, they discovered that random quantum circuits require specific gates set\citep{chamon_emergent_2014, shaffer_irreversibility_2014, zhou_single_2020} to generate entanglement complexity. 
This finding may be related to the information erasure mechanism in our study, though further research is required to elucidate the specific details.

To justify the ergodic hypothesis for equilibrium states in statistical mechanics through understanding the irreversibility of entropy increase in the second law of thermodynamics, we encounter several fundamental questions:
1. Conventionally, isolated systems with few particles exhibit periodic and reversible dynamics, whereas systems with large particle number $N$ and complex internal interactions demonstrate thermalization and irreversibility. What drives this transition? How does increasing particle number and interaction complexity transform the system's dynamics from periodic evolution to irreversible entropy increase? Does this transition occur continuously or through phase-transition-like abrupt changes?
2. During the evolution from non-equilibrium to equilibrium in isolated systems, the internal entanglement increases until reaching maximal entanglement at equilibrium, while subsystems (with particle numbers $\ll N$) become completely uncorrelated. How are inter-subsystem correlations erased during this process? Why does this lead to disentanglement failure and apparent irreversibility?
3. When an isolated system consists of two non-equilibrium subsystems with comparable sizes, it still thermalizes eventually. Does thermalization research necessarily require designating one large subsystem as an effective environment or heat bath?
To address these questions, in work \citep{guo_thermalization_2025} we constructed a gas-like model to study energy/particle transport and thermalization in non-equilibrium isolated systems. 
The model enables modification of internal interaction complexity by introducing additional particle types, adjusting particle numbers, and tuning interaction parameters.
The model captures real-world thermalization phenomena well.
Through numerical simulations, we demonstrated irreversible entropy increase and thermalization caused by erasure of spread non-equilibrium information. In the current work, we provide mathematical proofs explaining this entropy growth mechanism within our model's framework.

Boltzmann was the first to attempt developing a directional dynamical theory consistent with the second law of thermodynamics, hoping to show that any initial non-equilibrium distribution of gas would evolve to the final Maxwell velocity distribution. This research ultimately led to the discovery of the H-theorem. Boltzmann studied issues related to the irreversibility of entropy increase in the second law, but limited by classical mechanics concepts, he couldn't satisfactorily explain the origin of equilibrium states and the ergodic hypothesis (\textit{first proposed by Boltzmann and later formalized by Gibbs}).
After the establishment of quantum mechanics, the core ideas of Boltzmann's ergodic hypothesis required reinterpretation. When proposing the concept of the ergodic hypothesis for equilibrium states, Boltzmann relied on sampling either different subsystems or a single subsystem at different times, whereas Gibbs systematically formalized it through the concept of ensembles. Under the quantum mechanics framework, the ergodic hypothesis can be understood as equal-probability superpositions of all possible microstates, where the system always remains in a pure state.
Furthermore, in classical mechanics, the ergodic hypothesis considers all possible microscopic states in the phase space of an isolated system at a given energy $E$. The isolation requirement stems from energy conservation considerations—we must prevent any energy or particle leakage. \textit{In quantum mechanics, this stringent condition can be relaxed to the requirement of conserving the system's quasiparticle number.}
To erase the information of the non-equilibrium state, the application of the $\hat{O}$ operation slightly modifies both the energy expectation value and fluctuations of the system. While this does not strictly satisfy the isolation condition, it remains physically acceptable. \textit{These $\hat{O}$ operations can be set to conserve the system's quasiparticle number while influencing the system in an adiabatic manner.}  Moreover, this local operation $\hat{O}$ can be incorporated into isolated systems \citep{guo_thermalization_2025}.

In recent decades, some theoretical works studying system thermalization from a dynamical perspective include \citep{tasaki_quantum_1998, scarani_thermalizing_2002, scarani_entanglement_2007, linden_quantum_2009, short_equilibration_2011, short_quantum_2012, wilming_towards_2017, goldstein_extremely_2015, tasaki_typicality_2016,  rigol_relaxation_2007, manmana_strongly_2007, rigol_thermalization_2008, rigol_quantum_2009, rigol_quantum_2010, cassidy_generalized_2011, santos_entropy_2011, brenes_eigenstate_2020, huang_quantum_2025, harrow_thermalization_2023, shiraishi_systematic_2017, mori_thermalization_2017,roos_macroscopic_2025, tasaki_macroscopic_2024, shiraishi_nature_2024, tasaki_macroscopic_2024_2,teufel_typical_2024,vogel_long-time_2024}.
Among them, works \citep{tasaki_quantum_1998, scarani_thermalizing_2002, scarani_entanglement_2007, linden_quantum_2009, short_equilibration_2011, short_quantum_2012} adopted research approaches similar to \citep{popescu_entanglement_2006}, by dividing an isolated system into two parts, treating one smaller subsystem as the study object and the other part as an effective thermal bath, to investigate how the smaller subsystem's dynamical evolution leads to the canonical statistical distribution function.
Works \citep{tasaki_typicality_2016} studied the typicality of thermalization dynamics in isolated systems. Works \citep{rigol_relaxation_2007, manmana_strongly_2007, rigol_thermalization_2008, rigol_quantum_2009, rigol_quantum_2010, cassidy_generalized_2011, santos_entropy_2011, brenes_eigenstate_2020} used exact diagonalization to study chaotic properties of many-body systems and relaxation of observables. These results \citep{rigol_relaxation_2007, manmana_strongly_2007, rigol_thermalization_2008, rigol_quantum_2009, rigol_quantum_2010, cassidy_generalized_2011, santos_entropy_2011, brenes_eigenstate_2020}  could be predicted by the Generalized Gibbs Ensemble \citep{rigol_relaxation_2007} and understood through the eigenstate thermalization hypothesis \citep{rigol_thermalization_2008}.
Work~\citep{huang_quantum_2025} gave a proof of entropy thermalization in a particular quantum system.
Recent studies \citep{roos_macroscopic_2025,tasaki_macroscopic_2024, shiraishi_nature_2024} have investigated the thermalization of free fermion gases under perturbations, with related work also reported in \citep{tasaki_macroscopic_2024_2}.

We aim to prove the following content. Suppose a system is initially in an arbitrary non-equilibrium state, where the non-equilibrium can manifest as energy or particle distribution imbalances between two subsystems. During subsequent evolution, as energy and particles transfer and collide, the system state tends toward equilibrium, accompanied by increasing internal entanglement.
If we then perform 'information erasure' on particles in selected regions and implement reverse evolution, the entanglement generated during forward evolution cannot be completely undone. The final quantum state will be closer to equilibrium compared to the initial state.
Here we should clarify: when starting from a non-equilibrium state, the system may not always evolve directly toward equilibrium; it might first evolve to more non-equilibrium states. However, in such cases, the system inevitably enters periodic oscillations between non-equilibrium and 'equilibrium' states.
During these oscillations, when the system is in an 'equilibrium' state, information erasure will reduce the amplitude of return to non-equilibrium states. Eventually, the oscillation amplitude decays, driving the system toward equilibrium. This demonstrates that information erasure occurs repeatedly during the evolution from non-equilibrium to equilibrium states.
Across different thermalizable many-body models (non-integrable many-body systems), we can always identify certain subsystems that act as 'erasers', repeatedly eliminating spread non-equilibrium state information during system evolution.

We adopt the model from \citep{guo_thermalization_2025}, which assumes a lattice system containing multiple fermion types. For the same fermion type, each lattice site can accommodate only one particle, satisfying the Pauli exclusion principle, while different fermion types on the same site experience repulsive or attractive interactions. Here we consider a system with two fermion types $\tau$ and $\upsilon$, with particle numbers $N_\tau$ and $N_\upsilon$ respectively. Taking any non-equilibrium quantum state as the initial state $|\phi_0\rangle$, and denoting the possible Fock state basis vectors for each particle type as $\mathbf{x}$ and $\mathbf{y}$ with dimensions $d_x$ and $d_y$ respectively, the initial non-equilibrium state can be expressed as:
\begin{align}
|\phi_0\rangle &= \sum_{m=1}^{d_x}\alpha_{m,0}|x_m\rangle|\psi_{\upsilon,m,0}\rangle \label{phi0_1} \\
& = \sum_{n=1}^{d_y}\beta_{n,0}|y_n\rangle|\psi_{\tau,n,0}\rangle \label{phi0_2} \\
& = \sum_{m=1}^{d_x}\sum_{n=1}^{d_y}\gamma_{m,n,0}|x_m\rangle|y_n\rangle \label{phi0_3} .
\end{align}
The subscripts $m$ ($n$) index the $m$-th  ($n$-th) component of $\mathbf{x}$ ($\mathbf{y}$), $\tau$ and $\upsilon$ label different particle types, and $0$ indicates the time point corresponding to the quantum state. Since the system starts in a non-equilibrium state, the coefficients $\alpha_{i,0}$ ($\beta_{j,0}$) are non-uniformly distributed over the Fock state basis $\mathbf{x} = \left\{ |x_i\rangle \right\} _ {i=1}^{d_x}$ ($\mathbf{y} = \left\{ |y_i\rangle \right\}_{i=1}^{d_y}$). Similarly, $|\psi_{\tau,j,0}\rangle$ and $|\psi_{\upsilon,i,0}\rangle$ also show non-uniform distributions. We quantify the equilibrium properties using:
\begin{itemize}
\item $\tau$-subsystem: $S_{\tau} = -\sum_{m=1}^{d_x}p_{m,0}\ln p_{m,0}$
\item $\upsilon$-subsystem: $S_{\upsilon} = -\sum_{n=1}^{d_y}p_{n,0}\ln p_{n,0}$
\item Total system: $S = -\sum_{m=1}^{d_x}\sum_{n=1}^{d_y}p_{m,n,0}\ln p_{m,n,0}$
\end{itemize}
where $p_{m,0}=\alpha^*_{m,0}\alpha_{m,0}$, $p_{n,0}=\beta^*_{n,0}\beta_{n,0}$, and $p_{m,n,0} = \gamma^*_{m,n,0}\gamma_{m,n,0}$.
The system Hamiltonian is given by:
\begin{align}\label{hamiltonian}
\hat{H} &= \sum_{\langle i,j \rangle} \left[ J_\tau \left(c_{i,\tau}^\dagger c_{j,\tau} + c_{i,\tau} c_{j,\tau}^\dagger \right) + J_\upsilon \left(c_{i,\upsilon}^\dagger c_{j,\upsilon} + c_{i,\upsilon} c_{j,\upsilon}^\dagger \right) \right] \notag \\
&\quad + \sum_{i} [U_{i, \tau} n_{i,\tau} + U_{i, \upsilon} n_{i,\upsilon}] \notag \\
&\quad + \sum_{i} U_{\tau,\upsilon} n_{i,\tau} n_{i,\upsilon}.
\end{align}
Here we set the potential fields $U_{i, \tau} = U_{i, \upsilon} = 0$, with $U_{\tau,\upsilon}$ comparable to $J_\tau$ and $J_\upsilon$. This ensures neither the potential fields nor inter-particle interactions constrain particle mobility on the lattice.

Directly considering the system's time evolution is complex. Here, we adopt a stepwise evolution method to approximate the system's dynamics under the Hamiltonian $\hat{H}$. The entire evolution process is divided into two steps, with corresponding Hamiltonians $\hat{H}_1$ and $\hat{H}_2$ defined as:
\begin{align}\label{h1}
\hat{H}_1 &= \sum_{\langle i,j \rangle} J_\tau \left(c_{i,\tau}^\dagger c_{j,\tau} + c_{i,\tau} c_{j,\tau}^\dagger \right) \notag \\
&\quad + \sum_{i} U_{i, \tau} n_{i,\tau} \notag \\
&\quad + \sum_{i} U_{\tau,\upsilon} n_{i,\tau} n_{i,\upsilon}.
\end{align}
\begin{align}\label{h2}
\hat{H}_2 &= \sum_{\langle i,j \rangle} J_\upsilon \left(c_{i,\upsilon}^\dagger c_{j,\upsilon} + c_{i,\upsilon} c_{j,\upsilon}^\dagger \right) \notag \\
&\quad + \sum_{i} U_{i, \upsilon} n_{i,\upsilon} \notag \\
&\quad + \sum_{i} U_{\tau,\upsilon} n_{i,\tau} n_{i,\upsilon}.
\end{align}
In the first step, we evolve the initial wavefunction \eqref{phi0_2} under $\hat{H}_1$. During this process, only $\tau$ particles can move within the lattice while $\upsilon$ particles remain fixed. The Fock state $|y_n\rangle$ of $\upsilon$ particles provides an effective lattice potential for $\tau$ particles, superimposed on their original potential.
The $\tau$-particle wavefunction $|\psi_{\tau,n,0}\rangle$ evolves under the effective potential corresponding to the $\upsilon$-particle state $|y_n\rangle$, transitioning from the initial non-equilibrium state to a more balanced state $|\psi_{\tau,n,1}\rangle$. The system's wavefunction at this stage is denoted as:
\begin{align}
|\phi_1\rangle &= \sum_{m=1}^{d_x}\alpha_{m,1}|x_m\rangle|\psi_{\upsilon,m,1}\rangle \label{phi1_1} \\
& = \sum_{n=1}^{d_y}\beta_{n,1}|y_n\rangle|\psi_{\tau,n,1}\rangle \label{phi1_2} .
\end{align}
In this wavefunction, $\beta_{n,1} = \beta_{n,0}$ (the $\upsilon$-particle distribution remains unchanged), while the $\tau$-particle state $|\psi_{\tau,n,1}\rangle$ becomes more balanced compared to $|\psi_{\tau,n,0}\rangle$.
This evolution process has also been studied by Zurek \citep{zurek_pointer_1981,zurek_decoherence_2003,zurek_quantum_2018}.

In the second step, we evolve the system's wavefunction \eqref{phi1_1} under $\hat{H}_2$. During this process, $\tau$ particles remain fixed while $\upsilon$ particles move within the lattice. The Fock state $|x_m\rangle$ of $\tau$ particles provides an effective potential for $\upsilon$ particles, which is superimposed on their original potential.
The $\upsilon$-particle wavefunction $|\psi_{\upsilon,m,1}\rangle$evolves under the effective potential corresponding to the $\tau$-particle state $|x_m\rangle$, transitioning from the initial non-equilibrium state to a more balanced state $|\psi_{\upsilon,m,2}\rangle$. The system's wavefunction at this stage is denoted as:
\begin{align}
|\phi_2\rangle &= \sum_{m=1}^{d_x}\alpha_{m,2}|x_m\rangle|\psi_{\upsilon,m,2}\rangle \label{phi2_1} \\
& = \sum_{n=1}^{d_y}\beta_{n,2}|y_n\rangle|\psi_{\tau,n,2}\rangle \label{phi2_2} .
\end{align}
In this wavefunction, $\alpha_{m,2} = \alpha_{m,1}$ (the $\tau$-particle distribution remains unchanged), while the $\upsilon$-particle state $|\psi_{\upsilon,m,2}\rangle$ becomes more balanced compared to $|\psi_{\upsilon,m,1}\rangle$.

After the two-step evolution, both $\tau$ and $\upsilon$ particles reach more uniformly distributed states. At this stage, we perform a local information erasure operation $\hat{O}$. In \citep{guo_thermalization_2025}, the $\hat{O}$ operation acts on one lattice site, which is equivalent to multiplying the relevant quantum states by a phase factor. Here, we assume the application of a local operation $\hat{O}'$ that multiplies the Fock states of $\upsilon$ particles by random phase factors $e^{i\theta_{\upsilon,n}}$, transforming the state $|\phi_2\rangle$ into $|\phi'_2\rangle$:
\begin{align}
|\phi'_2\rangle &= \sum_{m=1}^{d_x}\alpha_{m,2}|x_m\rangle|\psi'_{\upsilon,m,2}\rangle \label{phi2'1} \\
& = \sum_{n=1}^{d_y}e^{i\theta{\upsilon,n}}\beta_{n,2}|y_n\rangle|\psi_{\tau,n,2}\rangle \label{phi2'_2} .
\end{align}

We now let the system's wavefunction \eqref{phi2'1} undergo reverse evolution under $-\hat{H}_2$ for an equal duration. Without the $\hat{O}'$ operation, the equilibrium wavefunction $|\psi_{\upsilon,m,2}\rangle$ would return to the non-equilibrium state $|\psi_{\upsilon,m,1}\rangle$.
Since $|\psi_{\upsilon,m,2}\rangle$ is more balanced than $|\psi_{\upsilon,m,1}\rangle$, we can intuitively consider $|\psi_{\upsilon,m,2}\rangle$ as having more non-zero components than $|\psi_{\upsilon,m,1}\rangle$. This implies that during the evolution from $|\psi_{\upsilon,m,2}\rangle$ to $|\psi_{\upsilon,m,1}\rangle$, both destructive and constructive interference must occur.
However, the current wavefunction $|\psi'_{\upsilon,m,2}\rangle$ has been multiplied by random phase factors, which disrupts the original interference patterns (both destructive and constructive) in the reverse process. Consequently, after the reverse evolution under $-\hat{H}_2$ for equal time, the $\upsilon$ particles remain in a more balanced state. We denote this quantum state as:
\begin{align}
|\phi'_1\rangle &= \sum_{m=1}^{d_x}\alpha_{m,1}|x_m\rangle|\psi'_{\upsilon,m,1}\rangle \label{phi1'1} \\
& = \sum_{n=1}^{d_y}\beta'_{n,1}|y_n\rangle|\psi'_{\tau,n,1}\rangle \label{phi1'_2} .
\end{align}
In this wavefunction, $\alpha_{m,1} = \alpha_{m,2}$ (the $\tau$-particle distribution remains unchanged), while the $\upsilon$-particle state $|\psi'_{\upsilon,m,1}\rangle$ becomes more balanced compared to $|\psi_{\upsilon,m,1}\rangle$.

Building upon this, we let the system's wavefunction \eqref{phi1'_2} undergo reverse evolution under $-\hat{H}_1$ for an equal duration. Since the $\upsilon$-particle wavefunction coefficients $\beta'_{n,1}$ are more balanced compared to the previous $\beta_{n,1}$, this means more non-zero $|y_n\rangle$ components participate in the evolution for the system's wavefunction.
These $\upsilon$-particle Fock state components $|y_n\rangle$ generate corresponding effective potential fields, under which the $\tau$-particle wavefunctions $|\psi'_{\tau,n,1}\rangle$ evolve and interfere.
If the original wavefunction \eqref{phi1_2} allowed the $|\psi_{\tau,n,1}\rangle$ components to evolve under their respective $\beta_{n,1}|y_n\rangle$ effective potentials and produce effective destructive and constructive interference that returned the system to its initial non-equilibrium state, then the new wavefunction \eqref{phi1'_2} will have this interference pattern disrupted during the evolution of $|\psi'_{\tau,n,1}\rangle$ under $\beta'_{n,1}|y_n\rangle$, preventing the $\tau$ particles from returning to their original non-equilibrium state.
The system's final quantum state is expressed as:
\begin{align}
|\phi'_0\rangle &= \sum_{m=1}^{d_x}\alpha'_{m,0}|x_m\rangle|\psi'_{\upsilon,m,0}\rangle \label{phi0'1} \\
& = \sum_{n=1}^{d_y}\beta'_{n,0}|y_n\rangle|\psi'_{\tau,n,0}\rangle \label{phi0'2} \\
& = \sum_{m=1}^{d_x}\sum_{n=1}^{d_y}\gamma'_{m,n,0}|x_m\rangle|y_n\rangle \label{phi0'_3} .
\end{align}
In this wavefunction, $\beta'_{n,0} = \beta'_{n,1}$. The state $|\phi'_0\rangle$ is more balanced compared to $|\phi_0\rangle$, meaning $S'_\tau > S_\tau$, $S'_\upsilon > S_\upsilon$, and $S' > S$.
In typical thermalization processes, the above procedure repeats until maximum entropy is reached. 
This information erasure process is represented by sequence of $\hat{O_i}(t_i)$. In actual thermalization processes, both $\hat{O_i}$ and $t_i$ can differ across iterations.

The second law of thermodynamics occupies a unique position in physics. This law fundamentally contradicts mechanical principles, and the conflict between them has long resisted reconciliation. Yet crucially, the Second Law serves as the foundation for equilibrium states and the ergodic hypothesis, which in turn constitute the basis of statistical physics - thus its paramount importance needs no elaboration.
Historically, research in statistical physics (represented by Boltzmann, who firmly believed in atomic hypothesis and dedicated himself to microscopic and statistical explanations of macroscopic phenomena) and quantum physics research (represented by Planck, the discoverer of energy quantization) are deeply entangled. The development of statistical physics directly contributed to the birth of quantum physics, strongly suggesting a profound underlying connection between these theories.
Our work demonstrates that the Second Law of Thermodynamics can be fundamentally understood and proven through the quantum physics and quantum information theory. This establishes quantum mechanics as the first-principles foundation of statistical physics, revealing the essential unity between these two theoretical frameworks.  Boltzmann's ergodic hypothesis emerges necessarily as an inevitable consequence of quantum dynamical evolution under well-defined conditions.

%

\bibliography{irreversibility2}

\end{document}